\documentstyle[twoside,fleqn,espcrc2,epsf]{article}

    \newcommand{\ba}{\begin{eqnarray}}
    \newcommand{\ea}{\end{eqnarray}}
    \newcommand{\be}{\begin{equation}}
    \newcommand{\ee}{\end{equation}}

\newcommand{\AmS}{{\protect\the\textfont2
  A\kern-.1667em\lower.5ex\hbox{M}\kern-.125emS}}

\title{Lattice Fermions and Chiral Symmetry}

\author{
        H. Banerjee
       \address{S.N. Bose National Centre for Basic Sciences, 
       JD Block, Sector III, Salt Lake, Calcutta 700091, India}
          and 
        Asit K. De 
        \address{Saha Institute of Nuclear Physics,          
        AF Block, Sector I, Salt Lake, Calcutta 700064, India}%
        \thanks{Speaker at the conference.}
         }
       
\begin{document}

\begin{abstract}
We propose a formulation of lattice fermions with one-sided 
differences that is hermitian, chirally symmetric (barring a bare mass term)
and completely free of doubling.
To obtain the axial anomaly in perturbation theory  
it was necessary to break chiral symmetry on the lattice only through
a bare mass term for the physical fermion. The chiral limit may be taken
once the continuum limit is reached. We comment on the role of the mass term
with examples elsewhere in field theory.
\end{abstract}

\maketitle

\section{Introduction}

The so-called `naive' lattice fermions give rise to sixteen species and the
absence of the axial anomaly in this case 
is well understood in terms of cancellation
of contributions from different species \cite{KaSm}.
The Wilson fermions avoid species `doubling' by introducing generalized
(momentum-dependent) mass terms through an irrelevant term that makes
the doublers as heavy as the cutoff and the axial anomaly comes out correctly
with the irrelevant Wilson mass term playing the role of a Pauli-Villars
regulator mass \cite{KeSe}; while a naive analysis gives fifteen times
the `conformal' anomaly \cite{Fuji}.
Apart from removing doublers, we will be concerned in the following 
with Ward 
identities in the continuum limit and the role of the bare mass term
and irrelevant terms in deriving the chiral anomaly.

\section{Hermitian Action with One-sided Differences}

All earlier attempts with
one-sided difference operators \cite{OSDF} to remove doublers
lacked hermiticity and as a result may have problems with reality properties
in the continuum limit and also with
numerical simulations. Most importantly, 
the issue of chiral symmetry breaking and the chiral 
anomaly was not addressed.

It is important
for anomaly-related issues \cite{Alva}
to keep intact, in the chiral
limit, the chiral and hermiticity structure of the Dirac
operator, 
$i\,{\cal D}\!\!\!\!/$. Accordingly,
we propose to use, in Weyl basis, for free fermions:
\begin{equation}
\label{eq:osdf}
i\,{\cal D}\!\!\!\!/ 
= \left(\begin{array}{cc}
        0 & D \\
        D^{\dagger} & 0 \end{array}\right) 
\equiv \left(\begin{array}{cc}
        0 & i\, \sigma_\mu \Delta_\mu^f \\
        i\, \sigma_\mu^\dagger \Delta_\mu^b & 0 \end{array}\right), 
\end{equation}
\begin{equation}
\mbox{where,}\,\,\,\Delta_{\mu}^f =  (\delta_{y,x+\mu} - \delta_{y,x})/a,
\end{equation}
\begin{equation}
\Delta_\mu^b  =  (\delta_{y,x} - \delta_{y,x-\mu})/a  
              =  - {(\Delta_\mu^f)}^\dagger \, ,
\end{equation}
and $\sigma_\mu =                                         
(i, \, \sigma_k)$ with $\sigma_k$ the usual Pauli matrices, $k=1,2,3$.
We shall see later the need to add, in the lattice-regularized theory, 
an explicit chiral symmetry breaking bare mass term.  

For massive Dirac fermions, the Dirac operator follows from 
Eq.~(\ref{eq:osdf}):
\begin{equation}
i\,{\cal D}\!\!\!\!/  = i\, \gamma_\mu \Delta_\mu^s
                        + i\, \gamma_\mu \gamma_5 \Delta_\mu^a + m \, ,
\end{equation}
\begin{equation}
\mbox{where}\,\,\, {\Delta}_{\mu}^s(x,y) = \frac{1}{2a} (\delta_{y,x+\mu}
                                         -\delta_{y,x-\mu}), 
\end{equation}
\begin{equation}
\mbox{and}\,\,\, \Delta_\mu^a(x,y) =  
\frac{1}{2a} (2 \delta_{y,x} - \delta_{y,x+\mu}
                                             - \delta_{y,x-\mu}).
\end{equation}
$\Delta_\mu^a$ produces an (irrelevant) term in 
the lattice action that formally goes to zero as $ a \rightarrow 0 $.

In a vectorlike gauge theory, 
the fermionic part of the euclidean lattice action 
is then given by,
\begin{eqnarray}
& S_F & = \sum_{x,\mu} \frac{1}{2a} \bar{\psi}_x \gamma_\mu 
      [U_{x\mu}\psi_{x+\mu}-U_{x-\mu,\mu}^\dagger\psi_{x-\mu}] \nonumber  \\
    & + & \sum_{x,\mu} \frac{1}{2a}\bar{\psi}_x \gamma_\mu \gamma_5
    [2\psi_x - U_{x\mu}\psi_{x+\mu} \nonumber \\ 
    & - & U_{x-\mu,\mu}^\dagger\psi_{x-\mu}] 
    + m \sum_x \bar{\psi}_x \psi_x \, , \label{eq:theory}
\end{eqnarray}
where $\psi_x$ and $\bar{\psi}_x$ are fermion fields at $x$ and 
$U_{x\mu}$ is the gauge field at the link $(x,\mu)$.
The action has full chiral symmetry for $m=0$.

The action (\ref{eq:theory}) leads to the free fermion propagator $G_0$ 
given in momentum space by
\begin{equation}
G_0^{-1} = i\gamma_{\mu}s_\mu + \gamma_\mu \gamma_5 c_\mu +m
\end{equation}
where $s_\mu = (\sin ak_{\mu})/a$ and
$c_\mu = (1 - \cos ak_{\mu})/a$.
To see that it is free from doublers, consider
\begin{equation}
{(G_0 G^\dagger_0)}^{-1}=\sum_\mu(s_\mu^2+c_\mu^2)+m^2 
                         +2\sigma_{\mu\nu} \gamma_5 s_\mu c_\nu
\end{equation}
Note that a zero
of $G_0^{-1}$ for $m=0$ is necessarily a zero of $Tr{(G_0 G^\dagger_0)}^{-1}$. 
But the latter can vanish only when $k_\mu=0$ for all $\mu$ in the first 
Brillouin zone.
Our propagator is essentially
different from the previous cases \cite{OSDF} 
where the $\gamma_5$ in the irrelevant
term is absent and the inverse propagator is expressible in the form
$\gamma_\mu f_\mu$ or $\sigma_\mu f_\mu$ 
where $f_\mu$ is a trigonometric function of the
momentum and there are at least some extra non-covariant excitations.

The irrelevant term in our action, however, breaks hypercubic and 
reflection symmetries \cite{PePe}. Reflection Symmetry  
is a necessary condition for reflection positivity, one of the
assumptions of the Nielsen-Ninomiya theorem in the euclidean theory.

As a remedy we propose to
use instead the Dirac operator 
$\gamma_{\mu}\Delta^{s}_{\mu} +
\gamma^{\epsilon}_{\mu}\gamma_{5}\Delta^{a}_{\mu}$ 
with
$\gamma^{\epsilon}_{\mu} = \gamma_\mu \epsilon_\mu$ $(\epsilon_\mu = \pm 1)$
and correlation functions
are to be evaluated only after averaging over $\epsilon_\mu$.
In Weyl language, this prescrition resolves the ambiguity, whether to take
the forward or the 
backward difference operator for a particular Weyl component, by
averaging correlations functions 
over such possibilities for each $\mu$-component of the difference
operator while maintaining the (anti)-correlation of the finite differences
for $L$- and $R$-handed fermions and hence preserving 
the chiral and hermiticity structure (\ref{eq:osdf}). 

A similar averaging prescription
has also been applied previously to recover covariance in case of point-split 
regularization \cite{John}. 
In actuality, in our case the $\epsilon_\mu$ averaging may not be needed 
after all in the continuum limit since only
an irrelevant term is concerned.

\section{The Chiral anomaly}
The locally derived Ward identities for the theory (\ref{eq:theory}) 
corresponding to the global U(1) vector and axial 
symmetries, found in the usual way, are 
\begin{equation}
\label{eq:VWI}
\langle\Delta_\mu^b J^{+}_{\mu}(x)\rangle = \langle Y \rangle \,\, ,
\end{equation}
\begin{equation}
\label{eq:AVWI}
\langle\Delta_\mu^b J^{+}_{\mu 5}(x) \rangle  = 
2m\,\langle\bar{\psi}_x \gamma_5 \psi_x\rangle + \langle X \rangle \,\, , 
\end{equation}
\begin{equation}
\mbox{with,}\,\langle Y \rangle = 
\langle \Delta_\mu^b J^{\epsilon -}_{\mu 5}(x) \rangle,\,\,
\langle X \rangle = \langle\Delta_\mu^b J^{\epsilon -}_{\mu}(x) \rangle,
\end{equation}
\begin{equation}
J^{\pm}_{\mu}(x) =
\frac{1}{2}\left[\bar{\psi}_x\gamma_{\mu}U_{x\mu}\psi_{x+\mu} 
\pm \bar{\psi}_{x+\mu}\gamma_{\mu}U^{\dagger}_{x\mu}\psi_x\right],
\end{equation}
\begin{eqnarray}
J^{\pm}_{\mu 5}(x) & = &
\frac{1}{2}[\bar{\psi}_x\gamma_{\mu}\gamma_{5}
U_{x\mu}\psi_{x+\mu} \nonumber \\
& \pm & \bar{\psi}_{x+\mu}\gamma_{\mu}\gamma_{5}
U^{\dagger}_{x\mu}\psi_x].
\end{eqnarray}
$J_\mu^{\epsilon -}$ and $J_{\mu 5}^{\epsilon -}$ are obtained respectively
from $J_\mu^{-}$ and $J_{\mu 5}^{-}$ by replacing $\gamma_\mu$ by
$\gamma_\mu^\epsilon$ in the above expressions.
Note that the naive continuum limit of the currents 
$J_\mu^{+}(x)$ and $J_{\mu 5}^{+}(x)$ give respectively the expected vector
and axial vector currents in the continuum, while $J_{\mu}^{\epsilon -}$
and $J_{\mu 5}^{\epsilon -}$ go in the naive continuum limit to zero. 
We like to point out that the theory is fully defined by the action and 
the measure. The Ward identities simply follow from them. There is an
ambiguity for identifying the global currents and the only criterion is for 
them to assume the expected form in the continuum limit. This is true also
for the usual treatment of global currents with Wilson fermions.

In weak-coupling perturbation theory, we now calculate the
$\langle Y \rangle$ and $\langle X \rangle$ in the limit $a \rightarrow 0$.
We start from the
operator representations:
\begin{eqnarray}
\langle Y \rangle & = & Tr\,\gamma_{5}\langle x \vert\left(G_{F}R\!\!\!/\,^{\epsilon} +
R\!\!\!/\,^{\epsilon}G_{F}\right)\vert x \rangle \label{eq:Y} \\
\langle X \rangle & = & Tr\,\langle x \vert\left(G_{F}R\!\!\!/\,^{\epsilon} -
R\!\!\!/\,^{\epsilon}G_{F}\right)\vert x \rangle  \label{eq:X}
\end{eqnarray}
where $R\!\!\!/\,^{\epsilon} =
\gamma_{\lambda}\epsilon_{\lambda}R_{\lambda}$ with
\begin{equation}
R_{\lambda} = (2 - U_{\lambda}e^{ip_{\lambda}a} -
e^{-ip_{\lambda}a} U_{\lambda}^{\dagger})/(2a) \,\,. 
\end{equation}
The full fermion propagator $G_F$ is given by
\begin{equation}
G^{-1}_{F} = D\!\!\!\!/\, + R\!\!\!/\,^{\epsilon}\gamma_{5}+m \,\, ,
\end{equation}
\begin{equation}
\mbox{where}, \,\,\, D_{\lambda} = (U_{\lambda}e^{ip_{\lambda}a} -
e^{-ip_{\lambda}a} U^{\dagger}_{\lambda})/(2a) \,\, . 
\end{equation}
In (\ref{eq:Y}) and (\ref{eq:X}) 
`$Tr$' stands for trace over $\gamma$-matrices, and, in
non-abelian gauge theory, also over the symmetry matrices.

Two factors play key role in the calculation of $\langle X \rangle$ 
and $\langle Y \rangle$ : (i)
terms odd in $R_{\lambda}$ drop out because of $\epsilon_\mu$-averaging,
and (ii) in the `physical' sector of the loop momentum $0 \leq
\vert k_{\lambda}\vert \leq {\pi/2a}, R_{\lambda}$ is of $0(a)$ whereas in the
`doubler' sector ${\pi/2a} \leq \vert k_{\lambda}\vert \leq {\pi/a}$
it is of $0\left({1/a}\right)$ and behaves like the mass of a Pauli-Villars
regulator field. One finally obtains 
$\lim_{a\rightarrow 0} \langle Y \rangle = 0$ 
so that the U(1) vector current is conserved:
\begin{equation}
\langle\partial_{\mu}J_{\mu}(x)\rangle = 0. 
\end{equation}
However, $\langle X \rangle = $ \\ 
\begin{displaymath}
2i \, tr (F_{\lambda\rho}\tilde{F}_{\lambda\rho})
\int^{\frac{\pi}{a}}_{-\frac{\pi}{a}}\frac{d^{4}k}{(2\pi)^{4}}
\frac{\sum_{\lambda}c^{2}_{\lambda}\,\Pi_{\alpha}
\cos ak_{\alpha}}{[\sum_{\lambda}(s^{2}_{\lambda}
+c^{2}_{\lambda})+m^{2}]^{3}}
\end{displaymath} 
\begin{equation}
= (i/2\pi^{4})tr(F_{\lambda\rho}\tilde{F}_{\lambda\rho})
\sum^{4}_{\nu=1}(-1)^{\nu}\,\, {^{3}C_{\nu - 1}} I_{\nu}, 
\label{anomaly}
\end{equation} 
where $F_{\lambda \rho}$ and 
$\tilde{F}_{\lambda\rho}$ are the field tensor and its dual, 
`{\it tr}' stands
for trace over symmetry matrices, and
in the continuum limit $I_{\nu} = {\pi^{2}/2\nu}$. Thus the
model reproduces the anomalous Ward identity:
\begin{equation}
\langle \partial_{\mu} J_{\mu 5}(x) \rangle = 2m \langle\bar{\psi}_x \gamma_5 \psi_x \rangle
-\frac{i}{16\pi^{2}} tr
(F_{\lambda\rho}\tilde{F}_{\lambda\rho}). 
\label{eq:anwi}
\end{equation}
Note that the subscript $\nu$ of $I_{\nu}$ has the meaning of
the number of components of the loop momentum with support in the
doubler sector ${\pi/2a} \leq \vert k_{\mu}\vert \leq {\pi/a}$.
For finite `$m$' the contribution from $\nu = 0$ vanishes in the
continuum limit. If, however, $m$ is zero the latter would exactly
cancel the anomaly. For a nonvanishing anomaly it is,
therefore, essential that the continuum limit is taken first and the
chiral limit $m = 0$, if necessary, afterwards. 

In Eq.~(\ref{eq:anwi}), the first or explicit symmetry breaking term is
of course a direct consequence of the nonzero mass and the second or 
anomalous term results essentially from the square of the irrelevant
term of the action. However, the irrelevant term alone cannot produce
the anomaly unless $m \neq 0$. 

\section{Conclusions}
We have been able
to achieve a doubler-free hermitian lattice fermion action 
producing the chiral anomaly in the continuum limit. 
Barring a bare mass term which can only receive multiplicative
renormalization, the action is chirally symmetric and therefore may be
adaptable to chiral gauge theories.
The presence of the mass term on the regulator is crucial to get the 
chiral anomaly. The chiral limit may only be taken once the cutoff is 
removed. There is growing evidence in helicity-flip interactions in
QED and QCD \cite{flip} also suggesting that the role of mass terms
is more than just soft symmetry breaking. 
In Wilson fermions too, the anomaly appears in a very
similar way. Our analysis 
indicates that breaking chiral symmetry only in the physical sector may be
enough unlike in the Wilson case where chiral symmetry is broken also by
mass terms for the doublers.

\end{document}